\documentclass{emulateapj}
\usepackage{natbib}
\usepackage{epsfig}

\shorttitle{BL Lac Optical flux and polarization studies}
\shortauthors{Gaur et al.}

\begin{document}

\title{Anti-correlated Optical Flux and Polarization Variability in BL Lac }

\author{Haritma Gaur\altaffilmark{1,2}, Alok C.\ Gupta\altaffilmark{2}, Paul J.\ 
Wiita\altaffilmark{3}, 
Makoto Uemura\altaffilmark{4}, Ryosuke Itoh\altaffilmark{4} and Mahito Sasada\altaffilmark{4,5} }
\altaffiltext{1}{Inter-University Centre for Astronomy and Astrophysics (IUCAA), Ganeshkhind,
Pune-- 411 007, India; haritma@iucaa.ernet.in} 
\altaffiltext{2}{Aryabhatta Research Institute of Observational Sciences (ARIES), Manora Peak,
Nainital -- 263 129, India}
\altaffiltext{3}{Department of Physics,  The College of New Jersey, P.O.\ Box 7718, Ewing, NJ 08628-0718, USA}
\altaffiltext{4}{Hiroshima Astrophysical Science Center, Hiroshima University,
Kagamiyama 1-3-1, Higashi-Hirohsima 739-8526, Japan}
\altaffiltext{5}{Department of Astronomy, Graduate School of Science, Kyoto University,
Sakyo-ku, Kyoto 606-8502, Japan}

\begin{abstract}

We present the results of photometric (V band) and polarimetric observations of the blazar BL Lac
during 2008--2010 using TRISPEC attached to the KANATA 1.5-m telescope in Japan. The data reveal a 
great deal of variability ranging from days to months with detection of strong variations in 
fractional polarization. The V band flux strongly anti-correlates with the degree of polarization 
during the first of two observing seasons but not during the second. The direction of the 
electric vector, however, remained roughly constant during all our observations. 
These results are consistent with a model with at least two emission regions 
being present, with the more variable component having a polarization direction nearly 
perpendicular to that of the relatively quiescent region so that a rising flux can 
produce a decline in degree of polarization. We also computed models involving helical 
jet structures and single transverse shocks in jets and show that they might also be able 
to agree with the anti-correlations between flux and fractional polarization. 

\end{abstract}



\maketitle

\section{Introduction}

BL Lacertae, also known as 1ES 2200+420 ($z=0.0688 \pm 0.0002$; Miller \& Hawley 1977),
is the prototype of blazar class and is hosted by an elliptical
galaxy consisting of stellar population of about 0.7 Gyr age (Hyvonen et al.\ 2007). 
The spectral energy distribution of BL Lacertae peaks at optical/IR bands and is 
classified as a low frequency peaked BL Lac object
(LBL; Fossati et al.\ 1998). These objects are characterized by high degree of optical polarization
and rapid flux variability at all wavelengths. Their spectra are dominated by  
featureless non-thermal continuua. The approaching jet of BL Lac is within $6-10 ^\circ$ of our line 
of sight and has flow speed of 0.981-0.994$c$, or a Lorentz factor of 
7.0 $\pm$ 1.8 (Jorstad et al.\ 2005).
It has been one of the favorite targets of several multi-wavelength campaigns organized by the
Whole Earth Blazar Telescope
(Raiteri et al.\ 2009; Villata et al.\ 2009 and references therein).

Polarization observations at different wavelengths together with flux measurements offer valuable 
information in trying to understand the behavior of blazars and to model their jet physics. 
It is commonly thought that the emission mechanism in blazars in radio-through-optical bands 
is predominantly 
synchrotron radiation that originates in the jets.
 Polarimetric observations are required to investigate the magnetic field structures 
in the jets that is necessary to produce synchrotron radiation.
In the case of the basic shock-in-jet model, because of the strengthened ordering of the magnetic 
field in the shocked region, one can expect a positive correlation between flux and polarization, 
 i.e., an increase of polarization with an increase of brightness 
(Marsher \& Gear 1985; Hughes et al.\ 1985; Marscher et al.\ 1996; Hagen-Thorn et al.\ 2008). 
The shock front partially orders the turbulent magnetic field along the shock, which is usually considered to be 
roughly perpendicular to the jet direction though oblique shocks are often likely (Hughes 2005).
However, if any newly emitted blob of plasma produces an increase in the total flux but possesses
either a chaotic magnetic field or one misaligned with the large scale field then the reverse correlation is possible
(Hagen-Thorn et al.\ 2002; Jorstad et al.\ 2006).   
Expanding upon the shock-in-jet model, a modest change in jet direction will yield significant changes in 
both total and polarized fluxes that can be either correlated or anti-correlated, though substantial 
changes in the position angle of the polarized emission is common in this case (Gopal-Krishna \& Wiita 1992). 
An analysis of earlier polarimetry of BL Lac with sparse photometric measurements indicated that the polarized
flux was relatively constant even while the total flux varied considerably (Hagen-Thorn et al.\ 2002).
Previous authors (Moore et al.\ 1982,
Kikuchi et al.\ 1988; Sillanp{\"a}{\"a} et al.\ 1993) have found that the polarization vector often rotates  in 
BL Lac and the similar blazar OJ 287.  Marscher et al.\ (2008) observed very smooth 
rotations of the polarization position angle by over $200 ^\circ$ as the flux rose.
The observed rotation of polarization angle is very likely to be a signature of the disturbance causing the flare 
passing through a section of the jet that contains a helical magnetic field.
Recently,  Raiteri et al.\ (2013) observed an increase in optical degree 
of polarization in BL Lac when the flux was lower during their observations in 2011--2012.
In this Letter, we exhibit strong evidence for a clear anti-correlation between the optical flux and 
fractional polarization for BL Lac over the course of an observing season.
 Our photopolarimetric observations were done using the 
1.5-m ``Kanata'' telescope at the Higashi-Hiroshima Observatory during the period May 2008--Jan 2010. 
In Section 2, we present the observations and data reduction procedure. Section 3 provides the
results and analysis of the photopolarimetric observations of BL Lac, while Section
4 presents the discussion and conclusions.
\noindent
\begin{figure}
\centering
\includegraphics[width=0.53\textwidth]{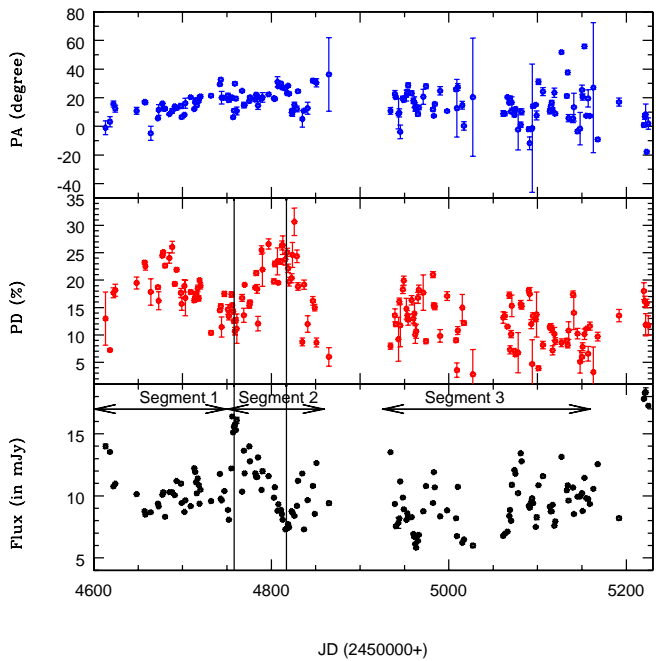}
\caption{The V passband light curve of the BL Lac over the two observing seasons 
during the period May 2008 -- Jan 2010 from KANATA telescope (lower panel), its percentage 
polarization  (middle panel) and polarization angle (in upper panel). Segments 1, 2 and 3 
are marked in the bottom panel of the figure.} 
\end{figure}

\begin{figure}
\centering
\includegraphics[width=0.52\textwidth]{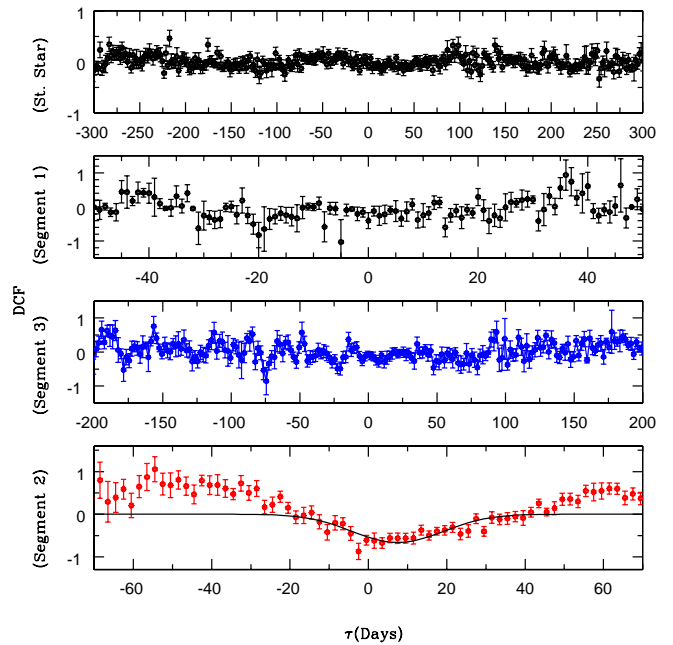}
\caption{ DCF for an unpolarized standard star is shown in the top panel. DCF between 
optical flux and polarization degree for the portions of the observations 
presented and marked in Fig.\ 1 as Segment 1, Segment 3  and Segment 2 are shown moving downwards; 
the solid line represents the fitted Gaussian function in Segment 2. }
\end{figure}

\section{Observations and Data Reduction}

We performed photopolarimetric observations of BL Lac
in the optical $V$ band over two observing seasons between May 2008-- Jan 2010. 
We divided the observations into three segments (shown in Fig.\ 1); the first  
consists of JD 2454613--4747, the second  is JD 2454751--4864 and the third is
JD 2454934--5224. We used TRISPEC attached to the 1.5-m
``Kanata'' telescope at Higashi-Hiroshima Observatory. TRISPEC is
capable of simultaneous three-band (one optical and two NIR bands)
imaging or spectroscopy, with or without polarimetry
(Watanabe 2005). 

The photometry of BL Lac was performed using standard CCD image reduction procedures.  
After making dark-subtracted and flat-fielded images, the magnitudes were measured using the aperture photometry technique. 
The radius of the aperture, which depended on the seeing each
night, was $\sim$3--5~arcsec. These correspond to 3--4 pixels on the optical
CCD.  
We calculated differential magnitudes of blazars using standard
stars located in the same frame. The standard stars are taken 
from Gonz{\'a}lez-P{\'e}rez et al. 2001.
We checked the constancy of the
brightness of the standard stars using the differential photometry
between them and neighboring stars in the same field.  

\begin{figure}
\centering
\includegraphics[width=0.52\textwidth]{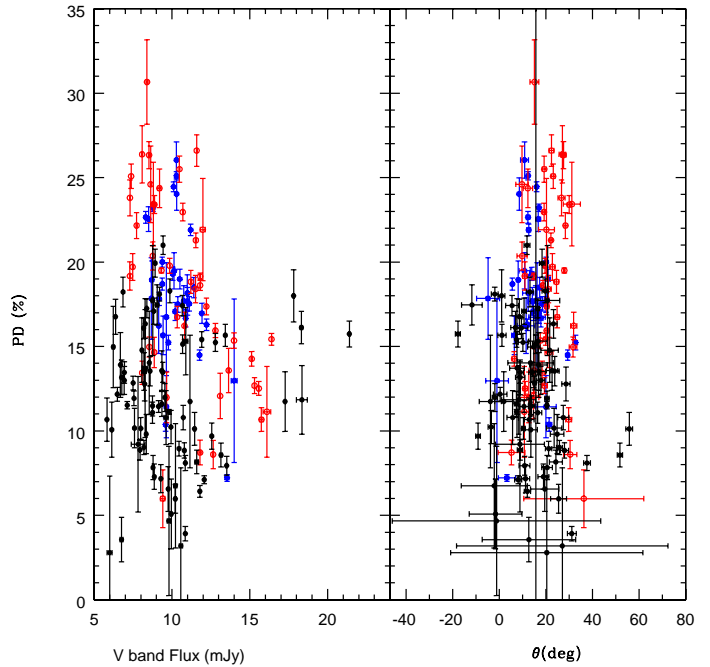}
\caption{The left panel shows the dependence between the degree of polarization and flux 
in the V band and the right panel shows the dependence between the degree of polarization 
and the polarization position angle. In both panels starred (blue) symbols represent 
segment 1, open (red) circles represent segment 2 and  solid (black) circles represent segment 3. 
The solid line in the right panel indicates the approximate direction of the parsec scale jet.}
\end{figure}

A set of polarization parameters was calculated from four consecutive
images, which were obtained with half-wave-plate angles of $0^\circ$,
$22.5^\circ$, $45.0^\circ$, and $67.5^\circ$. We took 12 sets of
images on one night, from which three sets of
polarimetric data were obtained.  Observations of unpolarized standard stars allowed us to
conclude that the instrumental polarization was less than 0.1\% in 
the $V$-band.   Hence we did not apply a
correction for instrumental polarization in obtaining the degree of fractional polarization ($PD$). 
The zero point of the polarization angle was
corrected to the standard system (measured from north to east).
Observations were sometimes carried out under bad sky
conditions. Some data obtained under such conditions have very large
errors and could disrupt any systematic trends that may exist in the
blazar's variability.  Therefore, in this paper we only use photometric data with
an error of less than 0.1~mag and $PD$ with an error less than 5\%. 
More details are given in Ikejiri et al.\ (2011).

\section{Analysis and Results}

In order to perform the cross-correlation between the optical flux and polarization, we have 
carried out Discrete Correlation Function (DCF) analyses. The DCF was first introduced by Edelson
\& Krolik (1988). For more details, see Hufnagel \& Bregman (1992), Hovatta et al.\ (2007) and
references therein.  

The photometric and polarimetric behavior of  BL Lac is shown in Fig.\ 1. The photometric
behavior is related with the polarization properties as optical fluxes apparently 
anti-correlate with the polarization levels,  strongly during the second half
of the first observing season. 
There is a gradual decline in the PD between the first and second observing seasons that
is likely related to long-term variations in the global
magnetic field. 

However, the polarization angle (PA) is roughly constant even in the most and least
active states of BL Lac we observed, with the PA rms error $\sim 7.144^\circ$.

Figure 1 is divided into three segments (shown by the arrows). Segment 1 is the first
half of the first observing season (JD 2454613--4747) and the DCF between flux and 
polarization degree is displayed in Fig.\ 2. We found  
no significant correlation between flux and polarization in this DCF.
Segment 3 consists of the entire second half of our data (JD 2454925--5160).
While there are multiple small peaks and dips in the DCF, both in flux and polarization, 
there is no significant positive or negative correlation present.  
Segment 2 is the second slice of the first observing season where the anti-correlation 
between optical flux and polarization degree appears most clearly. The strong dip present in 
the DCF is very close to a lag of 0, and this depression represents anti-correlation 
between flux and polarization. Around this negative peak the DCF curve is well fit by a 
Gaussian function of the form (shown in Fig.\ 2):
\begin{equation}
DCF(\tau)=A \times exp[\frac{-(\tau - m)^{2}}{2 \sigma^{2}}]
\end{equation}
where $A$ denotes the peak value of the DCF, $m$ represents the time lag in days at which the DCF peaks
and $\sigma$ represents the width of the Gaussian function. The values of these parameters are:
$A = -0.664; m=7.53 \pm 3.65$ and $\sigma =11.63 \pm$ 3.65.   The peaks found near $+60$ days
and $-55$ days are too close to the ends of the temporal period to ascribe physical meaning to them.
The variations in the DCF of the
unpolarized star also plotted in Fig.\ 2  show no significant peaks or dips.  This illustrates
that the wiggles seen in the BL Lac DCFs in Segments 1 and 3 are not significant and that no instrumental problem
was present during Segment 2.

Figure 3 illustrates the relation between the degree of polarization and V band flux 
for all the data.
Although there is no extremely strong trend, there also is an indication that the 
flux of BL Lac tended 
to be low when the PD was highest.  It is also clear from Fig.\ 3 that there is a significant
scatter in the PD when the flux is low, so when BL Lac is dimmer there are
low as well as high observed polarizations. Because there is an ambiguity of
 $\pm 180^{\circ}$ in the value of the PA,  we subtracted $180^{\circ}$ from the PAs 
to plot the dependence between PD and PA
shown in the right panel of Fig.\ 3.
The dependence between PD and PA possibly indicates that the polarization vectors are 
scattered around the dominant direction in the jet, which is shown by a straight line at 
$\sim$ $15.7^ \circ$ . This could be ascribed to a chaotic magnetic 
field being present in the emission region. Previous measurements
showed a different mean preferred polarization direction of $24^{\circ}$ (Hagen-Thorn et al.\ 2002).
and we have observed average PA of $15.7^{\circ}$.

\section{Discussion and Conclusions}

In most of the previous observations of blazars including polarimetry (e.g., Marscher et
al.\ 2008; Sasada et al.\ 2010; Marscher et al.\ 2010; Jorstad et al.\ 2010), a smooth
rotation of the polarization angle with the rise in optical flux has been noticed on
long term polarimetric observations. This can be explained by a non-axisymmetric magnetic
field distribution, a swing of the jet across our line of sight, or a curved trajectory
of the dissipation/emission pattern (Konigl \& Choudhuri 1985; Gopal-Krishna \& Wiita 1992;
Marscher et al.\ 2008).
 It also may be due to the propagation of a knot of emission that
follows a helical path in a magnetically dominated jet, as considered in the context of the
 event seen in BL Lac in long term observations in 2005--2006 (Marscher et al.\ 2008).
The large swings of polarization can be explained by ``bending jet" models where the angle
the jet makes with our line of sight varies (e.g., Gopal-Krishna \& Wiita 1992).
If variability arises from helical structures, the  observed polarization can be calculated 
following Lyutikov et al.\ (2005) and 
Raiteri et al.\ (2013).
The behavior of the observed polarization for optically thin synchrotron emission
with helical magnetic fields can be calculated using P=P$_{\rm max} \sin^2 \chi^{\prime}$,
where $\chi'$ is the viewing angle in the jet rest frame and is related to the 
observed viewing angle $\chi$ through the Lorentz transformation
\begin{equation}\sin \chi'= {{\sin \chi} \over {\Gamma_{\rm b} (1-\beta \cos \chi)}},
\end{equation}
where $\Gamma_{\rm b}$ is the bulk Lorentz factor of the plasma.

We compared the observed polarization with the polarization behaviour predicted by this
model. Following Raiteri et al.\ (2013), we varied  
$\Gamma_{\rm b}$ (7 $\pm$ 1.8, Jorstad et al.\ 2005) or $\chi_{\rm min}$ and found 
that when we lower $\Gamma_{\rm b}$   and $\chi_{min}$ 
(from Larionov et al.\ 2010) to 5.2 and 2$\degr$, respectively, in this case, the variations 
are more towards low polarization values (Fig.\ 4), but when we increase the values of 
$\Gamma_{\rm b}$ and $\chi_{max}$ (e.g. 8.8 and 6$\degr$ respectively in this case), the
variations are smaller and the  flux is lower but there is a higher degree of polarization.
From  Fig.\ 4, it is seen that the observed polarization of segment 2 can be very roughly 
reproduced by the first parameter set; however, the model always shows a flat 
trend in segment 3. 

\begin{figure}
\centering
\includegraphics[width=0.52\textwidth]{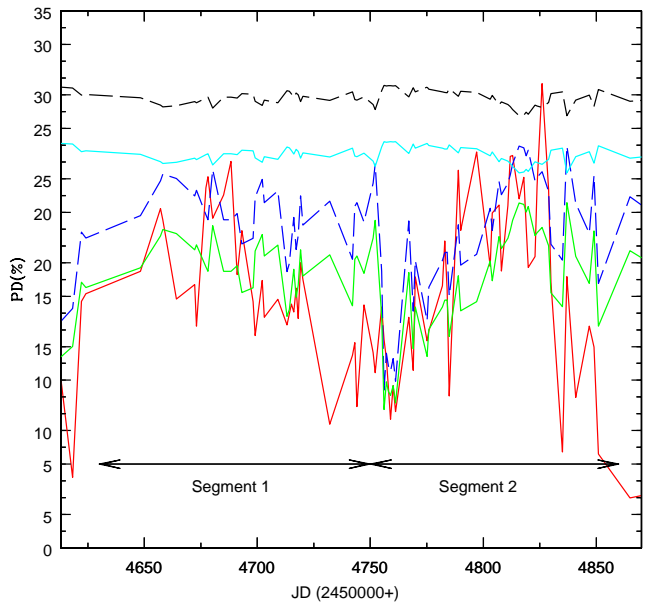}
\caption{The lower (red) plot shows the observed polarization curve, the dashed blue plot 
shows the polarization behaviour predicted by the helical magnetic 
field model (assuming $\Gamma_{\rm b}$ =5.2 and $\chi$=2$\degr$) and the dashed black plot 
shows the polarization behaviour by taking $\Gamma_{\rm b}$ = 8.8 and $\chi$=6$\degr$.
Similarly, the green plot shows the polarization behaviour predicted by the shock model
(assuming $\Gamma_{\rm b}$ =5.2 and $\chi$=2$\degr$) and cyan plot shows the polarization 
behaviour by taking $\Gamma_{\rm b}$ = 8.8 and $\chi$=6$\degr$. Only segments
1 and 2 are shown as the models always show a flat trend in segment 3 that consistently overpredicts 
the PA. } 
\end{figure}

The roughly constant PA seen in our observations is likely to arise
in a fairly uniform, straight and axially symmetric jet.  A perpendicular shock
moving along the jet which is viewed at a small but nearly constant angle to the jet axis
 would not result in a gradual change of polarization angle but it could lead to the
change in degree of polarization (Jorstad 2006).
If variability arises by the transverse shock wave model, the observed fractional polarization
of the shocked plasma radiation was calculated by Hughes et al.\ (1985) as
\begin{equation}
P \approx \frac{\alpha+1}{\alpha+5/3}~\frac{(1-k^{-2})\sin^2\chi'}{2-(1-k^{-2})\sin^2\chi'},
\end{equation}
where $(\alpha+1)/(\alpha+5/3)$ is the synchrotron polarisation factor due to a relativistic 
electron population with particle distribution ${\rm d}N/{\rm d}E \propto E^{-p}$, 
with $p=2 \, \alpha+1$, $k$ is the degree of compression of the shock wave and a value
$\sim$1.4 is chosen for the best agreement between observed and predicted polarization. 
Again we vary the values of $\Gamma_{\rm b}$ or $\chi_{\rm min}$ (similarly to the helical
jet model) and found that lowering the values of $\Gamma_{\rm b}$ or $\chi_{\rm min}$  
(and hence flux) would increase the variations at lower PD values, with the opposite being the case
for the higher values of $\Gamma_{\rm b}$ or $\chi_{\rm min}$ (these are shown in green 
and cyan color in fig 4). The shock model also seems to favor
the  set of parameters with smaller Lorentz factor and viewing angle, as the higher
values overpredict the PD.  
This model can also possibly explain the anti-correlation between flux and polarization.

The third possibility to explain the observations of strong anti-correlation 
between the flux and percentage polarization of BL Lac in segment 2 involves the existence of 
both an underlying, slowly varying component and short-lived variable components 
with different polarization directions (e.g., Hagen-Thorn et al.\ 2002; Uemura et al.\ 2010).
A specific scenario proposed by Marscher et al.\ (2008) has the radiation source 
consisting of two or more emission regions: one is a global jet region and the other are  local
emission regions. Here, the local emission arises from the highly polarized shocked ``clumps" 
moving inside the jet and is characterized by short-term variability because of the
small emission regions.  
The dependencies between flux level, PD and PA
seen in our data are consistent with this multi-component model. 
Polarization angles are clustered 
around the preferred direction (shown in Fig.\ 3) at the majority of flux levels; 
this reveals the presence of a constant underlying source always contributing to the total flux. 
Fig.\ 3 shows the inverse dependence of degree of polarization with respect to flux.
Newly formed polarized components lead to a rise in total flux but many randomly oriented
polarized components having comparable strengths but different position angles produce 
partial cancellations (e.g., Hagen-Thorn et al.\ 2002). 
This could lead to the observed decrease in total polarization at high brightness levels.
At low brightness levels (when only a few variable components contribute), there is significant 
scatter in the degree of polarization.  
Then the appearance of an additional highly polarized component with position angle 
along the preferred direction will increase the relative strength of the underlying source, 
thus resulting in higher observed polarizations at low flux levels.  
However, if this new polarized component has position angle perpendicular to the preferred 
direction, it can significantly cancel the polarization of the underlying source, 
causing low polarization to be observed. 
This can explain the significant scatter in the degree of polarization when the flux is low.

 We have used Monte Carlo simulations with two components to estimate the
probabilty of observing an anti-correlation between flux and polarization, and if it is present,
the probability of seeing PAs as constant as those actually observed. To do so, we obtained the underlying
probability distributions of the fluctuations of Stoke's parameters Q, U and I in the actual data. 
Using these probability distributions, we generated 
15,000 random realizations of season-long light curves (LCs) for flux and polarization 
degree as well as polarization angles 
with the underlying statistical properties of the original LCs for flux, PD and PA.
Next, taking these randomly generated LCs, we determined the 
DCFs between flux and polarization and found the probability of finding a strong anti-correlation 
to be 0.005 if we put 0.50 as the conservative minimum required DCF value (our actual
DCF had a maximum absolute value of 0.66). 
Finally, we calculated the probability of getting nearly constant polarization angles 
($\pm$ 20 degrees of the mean PA) out of the LCs showing anti-correlation between 
flux and polarization and found it to be $p=0.08$. Hence the chance of randomly producing a situation
similar to that we observed using this model is quite low and most of the time one
would expect to see either no correlation between flux and PD or a positive one, as is indeed usually seen.
The strong anti-correlation between flux and PD seen during a portion of the first year of our 
observations support the hypothesis that the local emission is very important in Segment 2,
while the global emission is dominant in Segments 1 and 3, indicating that the second component
became strong rather quickly and by the following
observing season this second component had faded.  This interpretation is strengthened as the anti-correlation appears after a flare with a very rapid rise-time.\\
Marscher et al.\ (2008) showed that the optical flux and polarization variability in BL Lac seen in 2005,
which included a large swing in the PA coincident with rapid changes in the PD,
is very nicely explained in terms of a shock wave leaving the vicinity of the central black hole and
 propagating down only a portion of the jet's cross section. In this case the disturbance follows a spiral path
in a jet that is both accelerating and becoming more collimated.
This interpretation is supported for that flare by the presence of a bright superluminal knot in their VLBA radio  maps
 and the agreement between the optical and 7mm radio polarization directions.
The relative constancy of the PA during our observations seems to indicate that this 
particular phenomenon was not being observed. 
However, recently Larinonov et al.\ (2013) have extended the Marscher et al.\ (2008) model to 
model multi-wavelength variations
of an outburst in the blazar  0716+714. They allowed
for variations in the bulk Lorentz factor, $\Gamma$, jet viewing angle, temporal evolution of the outburst, shocked
plasma compression ratio, $k$, spectral index $\alpha$, and pitch angle of the spiral motion.   
They found that even if all of these parameters, other than $\Gamma$, 
were fixed, a wide variety of flux and polarization
behaviors still could be reproduced (Larinonov et al.\ 2013).
We conclude that a temporary anti-correlation between total flux and PD,  even while the PA remains
nearly constant, such as  we found in BL Lac in late 2008, actually can be incorporated into the 
shock-in-spiral-jet picture, as long as the shock Lorentz factor is relatively low.
\\

HG thanks Archana Soam for discussions on polarization.  We acknowledge the referee's comments and suggestions 
which have substantially improved the manuscript.
A part of this work was supported by JSPS KAKENHI Grant Number 25120007
 and a Grant-in-Aid for JSPS Fellows.


\begin{thebibliography}{}

\bibitem[\protect\citeauthoryear{{Abdo} et~al.}{{Abdo}
  et~al.}{2009}]{abdo2009}
{Abdo}, A.~A., et al. 2009, \apj, 707, 1310




\bibitem[Edelson 
\& Krolik(1988)]{1988ApJ...333..646E} Edelson, R.~A., \& Krolik, J.~H.\ 1988, \apj, 333, 646 

\bibitem[Fossati et al.(1998)]{1998MNRAS.299..433F} Fossati, G., Maraschi, 
L., Celotti, A., Comastri, A., \& Ghisellini, G.\ 1998, \mnras, 299, 433 

\bibitem[Gonz{\'a}lez-P{\'e}rez et al.(2001)]{2001AJ....122.2055G} 
Gonz{\'a}lez-P{\'e}rez, J.~N., Kidger, M.~R., 
\& Mart{\'{\i}}n-Luis, F.\ 2001, \aj, 122, 2055 


\bibitem[Gopal-Krishna \& Wiita(1992)]{1992A&A...259..109G}
Gopal-Krishna, \& Wiita, P.~J. 1992, \aap, 259, 109

\bibitem[Hagen-Thorn et al.(2002)]{ 2002A&A...385...55H} Hagen-Thorn, V.~A., Larionva, E.~G., Jorstad, S.~G.,
Bj{\"o}rnsson, C.~I., Larionov, V.~M.,  2002, \aap, 385, 55

\bibitem[Hagen-Thorn et al.(2008)]{2008ApJ...672...40H} Hagen-Thorn, V.~A., 
Larionov, V.~M., Jorstad, S.~G., et al.\ 2008, \apj, 672, 40 

\bibitem[Hovatta et 
al.(2007)]{2007A&A...469..899H} Hovatta, T., Tornikoski, M., Lainela, M., et al.\ 2007, \aap, 469, 899 

\bibitem[Hufnagel 
\& Bregman(1992)]{1992ApJ...386..473H} Hufnagel, B.~R., \& Bregman, J.~N.\ 1992, \apj, 386, 473 

\bibitem[Hughes (2005)]{2005ApJ...621..635H} Hughes, P.~A., 2005, \apj, 621. 635
 
\bibitem[Hughes et al.(1985)]{1985ApJ...298..301H} Hughes, P.~A., Aller, 
H.~D., \& Aller, M.~F.\ 1985, \apj, 298, 301 


\bibitem[Hyv{\"o}nen et al.(2007)]{2007A&A...476..723H} Hyv{\"o}nen, T., Kotilainen, J.~K., Falomo, R., {\"O}rndahl, E., \& Pursimo, T.\ 2007, \aap, 476, 723 

\bibitem[Ikejiri et al.(2011)]{2011PASJ...63..639I} Ikejiri, Y., Uemura, 
M., Sasada, M., et al.\ 2011, \pasj, 63, 639 

\bibitem[Jorstad et al.(2005)]{2005AJ....130.1418J} Jorstad, S.~G., 
Marscher, A.~P., Lister, M.~L., et al.\ 2005, \aj, 130, 1418 

\bibitem[Jorstad et al.(2006)]{2006ChJAS...6a.247J} Jorstad, S., Marscher, 
A., Stevens, J., et al.\ 2006, Chinese Journal of Astronomy and 
Astrophysics Supplement, 6, 010000 

\bibitem[Jorstad et al.(2010)]{2010ApJ...715..362J} Jorstad, S.~G., 
Marscher, A.~P., Larionov, V.~M., et al.\ 2010, \apj, 715, 362 

\bibitem[Konigl 
\& Choudhuri(1985)]{1985ApJ...289..188K} Konigl, A., \& Choudhuri, A.~R.\ 1985, \apj, 289, 188 

\bibitem[Kikuchi et 
al.(1988)]{1988A&A...190L...8K} Kikuchi, S., Mikami, Y., Inoue, M., Tabara, H., \& Kato, T.\ 1988, \aap, 190, L8 

\bibitem[Larionov et al.(2013)]{2013ApJ...768...40L} Larionov, V.~M., Jorstad, S.~G., Marscher, A.~P. et al.
2013, \apj, 768, 40

\bibitem[Lyutikov et al.(2005)]{2005MNRAS.360..869L} Lyutikov, M., Pariev, 
V.~I., \& Gabuzda, D.~C.\ 2005, \mnras, 360, 869 

\bibitem[Marscher 
\& Gear(1985)]{1985ApJ...298..114M} Marscher, A.~P., \& Gear, W.~K.\ 1985, \apj, 298, 114 

\bibitem[Marscher(1996)]{1996ASPC..110..248M} Marscher, A.~P.\ 1996, Blazar 
Continuum Variability, 110, 248 


\bibitem[Marscher et al.(2008)]{2008Natur.452..966M} Marscher, A.~P., 
Jorstad, S.~G., D'Arcangelo, F.~D., et al.\ 2008, \nat, 452, 966 

\bibitem[Marscher et al.(2010)]{2010ApJ...710L.126M} Marscher, A.~P., 
Jorstad, S.~G., Larionov, V.~M., et al.\ 2010, \apjl, 710, L126 

\bibitem[Moore et al.(1982)]{1982ApJ...260..415M} Moore, R.~L., Angel, 
J.~R.~P., Duerr, R., et al.\ 1982, \apj, 260, 415 


\bibitem[Raiteri et 
al.(2009)]{2009A&A...507..769R} Raiteri, C.~M., Villata, M., Capetti, A., et al.\ 2009, \aap, 507, 769 

\bibitem[Raiteri et al.(2013)]{2013MNRAS.tmp.2378R} Raiteri, C.~M., 
Villata, M., D'Ammando, F., et al.\ 2013, \mnras, 2378 


\bibitem[Sasada et al.(2010)]{2010PASJ...62..645S} Sasada, M., Uemura, M., 
Arai, A., et al.\ 2010, \pasj, 62, 645 

\bibitem[Sillanp{\"a}{\"a} et 
al.(1993)]{1993Ap&SS.206...55S} Sillanp{\"a}{\"a}, A., Takalo, L.~O., Nilsson, K., \& Kikuchi, S.\ 1993, \apss, 206, 55 


\bibitem[Uemura et al.(2010)]{2010PASJ...62...69U} Uemura, M., Kawabata, 
K.~S., Sasada, M., et al.\ 2010, \pasj, 62, 69 




\bibitem[Villata et 
al.(2009)]{2009A&A...501..455V} Villata, M., Raiteri, C.~M., Larionov, V.~M., et al.\ 2009, \aap, 501, 455 

\bibitem[Watanabe et al.(2005)]{2005PASP..117..870W} Watanabe, M., Nakaya, 
H., Yamamuro, T., et al.\ 2005, \pasp, 117, 870 

\end{thebibliography}
\end{document}